\documentclass[12pt]{article} 

\usepackage{latexsym} % Gets \Box etc
\usepackage{amssymb}  % \gtrsim, \geqslant, etc etc: see amsguide.ps
\usepackage{amsbsy}   % \pmb and \boldsymbol
\usepackage{graphicx}       % For PostScript figures
\newcommand{\De}{\Delta}

\newcommand{\Om}{\Omega}

\newcommand{\p}{\partial}
\newcommand{\<}{\langle} 
\renewcommand{\>}{\rangle} % LaTeX: \> already defined

\newcommand{\txt}{\textstyle}

\newcommand{\dsp}{\displaystyle}

\newcommand{\beq}{\begin{equation}}
\newcommand{\eeq}{\end{equation}}
\newcommand{\ba}{\begin{array}}
\newcommand{\bea}{\begin{eqnarray}}
\newcommand{\ea}{\end{array}}
\newcommand{\eea}{\end{eqnarray}}
\newcommand{\bi}{\begin{itemize}}  %\setlength{\itemsep}{0\parsep}}
\newcommand{\ei}{\end{itemize}}
\newcommand{\ben}{\begin{enumerate}} %\setlength{\itemsep}{0\parsep}}
\newcommand{\een}{\end{enumerate}}
\newcommand{\bc}{\begin{center}}
\newcommand{\ec}{\end{center}}
\newcommand{\bl}{\begin{flushleft}}
\newcommand{\el}{\end{flushleft}}
\newcommand{\br}{\begin{flushright}}
\newcommand{\er}{\end{flushright}}

\newcommand\comment[1]{ \hbox{[{\it Comment suppressed here.}\/]} }
\newcommand\hide[1]{}

\newcommand{\skipover}[1]{}

\newcommand{\half} {{\txt \frac{1}{2}}}
\newcommand{\third}{{\txt \frac{1}{3}}}
\newcommand{\quarter}{{\txt\frac{1}{4}}}
\newcommand{\twothirds}{{\txt \frac{2}{3}}}
\newcommand{\sixth}{{\txt \frac{1}{6}}}

\newcommand{\threequarters}{{\txt \frac{3}{4}}}

\makeatletter %\catcode`\@=11

\def\appendix{\par                              % Have \appendix say
    \setcounter{section}{0}                     % `Appendix A', not just `A'
    \setcounter{subsection}{0}
    \renewcommand{\theequation}{\Alph{section}.\arabic{equation}}
    \renewcommand{\thesection}{Appendix \Alph{section}
                \setcounter{equation}{0}  } %Have eqns numbered (A.1) etc
}

\def\applabel#1{\@bsphack
  \protected@write\@auxout{}%
         {\string\newlabel{#1}{{\Alph{section}}{\thepage}}}%
  \@esphack}
\def\section{
\setcounter{equation}{0}        % Reset eqn numbers at start of section
\@startsection {section}{1}{\z@}{-3.5ex plus -1ex minus 
 -.2ex}{2.3ex plus .2ex}{\large\bf}}
\renewcommand{\theequation}{\arabic{section}.\arabic{equation}}

\def\subsection{\@startsection{subsection}{2}{\z@}{-3.25ex plus -1ex minus 
 -.2ex}{1.5ex plus .2ex}{\normalsize\bf}}

\def\subsubsection{\@startsection{subsubsection}{3}{\z@}{-3.25ex plus
 -1ex minus -.2ex}{1.5ex plus .2ex}{\normalsize}}

\makeatother   %\catcode`\@=12

\newsavebox{\eqlabel}
\makeatletter  %\catcode`\@=11
\newlength{\numblen}
\newsavebox{\eqnumb}
\def\@eqnnum{\savebox{\eqnumb}{\rm (\theequation)}%
\settowidth{\numblen}{\usebox{\eqnumb}}%
\makebox[\numblen][l]{\usebox{\eqnumb}~~~\usebox{\eqlabel}}}
\makeatother   %\catcode`\@=12

\newenvironment{equationwithlabel}[1]{ %
  \begin{equation}\label{#1} }{\end{equation}} %\savebox{\eqlabel}{~}}
\newcommand{\beql}[1]{\begin{equationwithlabel}{#1}}
\newcommand{\eeql}{\end{equationwithlabel}}
\newcommand{\MeV}{{\rm MeV}} 
 
\newcommand{\Qt}{{\tilde Q} }
\newcommand{\sbar}{{\bar s}}
\newcommand{\diag}{{\rm diag}}

\newcommand{\neut}{{\rm neutral}}
\newcommand{\unp}{{\rm unpaired}}

\begin{document}

\title{\bf Absence of two-flavor color superconductivity in compact stars}

\author{
Mark Alford${}^{(a)}$ and Krishna Rajagopal${}^{(b)}$
\\[0.5ex]
\parbox{0.6\textwidth}%
{
\normalsize
\begin{itemize}
\item[${}^{(a)}$] 
  Physics and Astronomy Department, \\
  Glasgow University, \\
  Glasgow~G12~8QQ, U.K.
\item[${}^{(b)}$] 
  Center for Theoretical Physics, \\
  Massachusetts Institute of Technology, \\
  Cambridge, MA~02139, U.S.A.
\end{itemize}
}}

\newcommand{\preprintno}{
  \normalsize GUTPA/02/03/02\ \ \ \ MIT-CTP-3258
}

\date{May 16, 2002 \\[1ex] \preprintno}

\begin{titlepage}
\maketitle
\def\thepage{}          % No page number on title page

\begin{abstract}
%       1         2         3         4         5         6
The simplest pattern of color superconductivity 
involves BCS pairing between  
up and down quarks. We argue that
this ``2SC'' phase 
will not arise within a compact star.
A macroscopic volume of quark matter must be electrically
neutral and must be a color singlet.  Satisfying these
requirements imposes a significant free energy cost on the 2SC phase,
but not on color-flavor locked (CFL) quark matter, 
in which up, down and strange
quarks all pair.  As a function of increasing density, therefore, one
may see a single phase transition from hadronic matter 
directly to CFL quark matter. Alternatively,
there may be an intervening phase
in which the different flavors self-pair, or
pair with each other in a non-BCS pattern,
such as in a crystalline color superconductor.
\end{abstract}

\end{titlepage}

\renewcommand{\thepage}{\arabic{page}}
%\setcounter{page}{1}

%--------------------------------------- 
                        % Body of paper begins
\section{Introduction}
\label{sec:intro}

We are beginning to learn some interesting things about
matter at very high densities. For experimental information, we have
to rely on observations of compact stars, which consist of
matter at nuclear density at the surface, increasing to
an unknown maximum density at the core. So far these observations
have not put strong constraints on the behavior of matter beyond 
nuclear density.
Theoretically, several different possible phases have been conjectured
for hyperdense matter of the kind we would expect to find in compact stars.
Some of the most interesting possibilities are
the color-superconducting phases expected if the density
is high enough to transform hadrons into quark 
matter~\cite{Barrois,BailinLove,ARW1,RappETC,CFL,Reviews}.
Color superconductivity occurs because
QCD predicts an attractive force between quarks that are 
antisymmetric in color, so we expect quarks
near their Fermi surfaces to form Cooper pairs, which condense,
breaking the color gauge symmetry.
One pairing pattern, known to be favored at sufficiently high density,
is the color-flavor locked (CFL) phase in which up-down, down-strange,
and up-strange Cooper pairs all form, allowing quarks of all three
colors and all three flavors to pair~\cite{CFL}.  At lower densities,
where the strange quark mass disfavors pairing of the strange with the 
light quarks,  previous work has suggested that the favored phase
is a form of color superconductivity
in which up and down quarks of two of the three colors participate in
pairing.  This ``2SC'' phase leaves the third color and strange
quarks unpaired.\footnote{
The quark matter that we discuss in this paper always contains strange quarks,
which may 
or may not pair with the light quarks. The phase where they do not,
which we here call ``2SC'', therefore contains unpaired strange quarks,
and is the phase we have called ``2SC+s'' elsewhere~\cite{ABR2+1}.}

In this paper we investigate what forms of superconducting quark matter
are most likely to occur in compact stars, taking into account the constraints
that must be imposed for a stable bulk phase.
These are:
\bi
\item Electromagnetic neutrality. 
Bulk matter must be neutral with respect to any Abelian gauge charge
(broken or unbroken). If it were not, the energy
would be nonextensive, growing faster than the volume as the size
of the sample increases.
\item Color neutrality. A macroscopic chunk of quark matter
must be a color singlet. Color neutrality, the equality in
the number density of red, green and blue color charges,
is a prerequisite for this.  Indeed, the authors
of Ref.~\cite{Amore} have shown that so long as a macroscopic
chunk of color superconductor is color neutral, carrying out
the projection which imposes color singlet-ness has negligible
effect on the free energy of the state.  This result is analogous
to the familiar fact from ordinary superconductivity that
the projection which turns a 
BCS state, wherein particle
number is formally indefinite, into a state with definite
but very large particle number has no significant effect.
Here, ``macroscopic'' means that the chunk of quark matter
must be much larger than the inverse of the
gap $\Delta$~\cite{Amore}, which is of order $10$ to $100$ MeV  
in a color superconductor~\cite{ARW1,RappETC,CFL,Reviews}.  
We are interested
in color superconducting regions of order kilometers in size.
Thus to very good approximation all we need to worry about
is color neutrality.  Color singlet-ness follows without
paying any further free energy price.
\ei

We shall see that when we impose these constraints,
the free energy of the 2SC phase becomes large enough that it
is unlikely to be found in nature.  
The CFL phase, by contrast, satisfies the neutrality
constraints at a much smaller cost. (The free energy
cost of imposing neutrality turns out to be less in the CFL phase than in
unpaired quark matter.)

Early work on the 2SC phase ignored issues of color and electromagnetic
neutrality, focussing instead on foundational issues like the size of
the pairing gap.  Calculations were performed in various toy models in
which QCD with only two flavors of quarks was analyzed with the
assumption of equal chemical potentials and equal number densities for
up and down quarks.
Subsequently, many authors have assumed that there
is some range of values of the strange quark mass for which
up-strange and down-strange pairing can be neglected, and so
a stable bulk 2SC phase would occur.  
We argue here that this is not the
case.  If the strange quark mass precludes
up-strange and down-strange pairing, then up-down pairing (i.e.~the 2SC phase)
is also precluded.  Other authors have looked at
the implications of maintaining electric and color neutrality
in the 2SC phase \cite{BaymIida}, 
but the very existence of stable bulk 2SC
quark matter has not previously been questioned.

In this paper we shall consider matter containing up ($u$), down ($d$)
and strange ($s$) quarks, but no heavier quarks. We will take the
$u$ and $d$ to be massless, and the strange quark to have an effective
(``constituent'') mass $M_s$, which may depend on the density and
phase, and is larger than its current mass $m_s\approx 120~\MeV$.
We shall assume that the quark number chemical potential
$\mu$ is large compared to $M_s$, and work only
to lowest nontrivial order in $M_s^2/\mu^2$.  This simplifies our
analysis considerably, but we should remark that it is unlikely
that $\mu$ is much larger than 500 MeV even in the
center of a compact star,
meaning that $M_s^2/\mu^2$ may not be very small.

\section{Enforcing neutrality}

We shall limit our discussion to quark matter in compact stars older
than a few minutes. This justifies our working at zero
temperature, since the temperature is by then well below
the expected value of all gaps and quasiparticle masses.
It also means that there has been plenty of time to
come to equilibrium under the weak interactions,\footnote{Other 
authors have considered color superconductivity
in contexts where the weak interactions
are not in equilibrium, as would for example
be appropriate if it were somehow possible to 
realize color superconductivity in a heavy ion collision
\cite{Bedaque:2002nu}.} 
so flavor symmetries are broken.
It also means that the star is transparent to neutrinos,
which leave the system, so lepton number is not conserved.  The
relevant symmetries are therefore just the color and electromagnetic
gauge symmetries and the global symmetry related to baryon number
conservation:
\beq
[SU(3)_{\rm color}]
 \times [U(1)_Q] 
 \times U(1)_B \ .
\eeq
In any color superconducting 
phase, this symmetry is broken down to some subgroup by condensation
of quark pairs.  

The baryon number of an isolated macroscopic body is fixed,
but since baryon number is not a gauge symmetry, nothing
prevents matter from having a constant nonzero baryon density
in the infinite volume limit.  There is no requirement
for ``baryon-number neutrality'', and we can treat the chemical potential
for baryon number, $3\mu$, as a free parameter.
A color superconductor is
a BCS state in which baryon number is
spontaneously broken by virtue of a condensate
wherein $\langle qq \rangle\neq 0$.  A macroscopic color
superconductor may
therefore seem to have ill-defined baryon number, but just
as in an ordinary superconductor or superfluid, and as shown explicitly
in Ref.~\cite{Amore}, projecting the BCS state onto a state
of fixed large baryon number has no significant effect.

Because we are only concerned with enforcing color neutrality, as
opposed to color singlet-ness, we need only 
consider the the $U(1)_3 \times
U(1)_8$ subgroup of the 
color gauge symmetry generated by the Cartan subalgebra $T_3 =
\diag(\half,-\half,0)$ and $T_8=\diag(\third,\third,-\twothirds)$ in
color space, where we choose to label the colors as $(r,g,b)$.  We
introduce chemical (color-electrostatic) potentials $\mu_3$ and
$\mu_8$ coupled to the color charges $T_3$ and $T_8$.  Enforcing color
neutrality means choosing $\mu_3$ and $\mu_8$ such that the $T_3$ and
$T_8$ densities vanish. Enforcing $T_3$ and $T_8$ neutrality suffices
to enforce equality in the number of red, green and blue quarks.

Electromagnetism is generated by 
$Q=\diag(\twothirds,-\third,-\third)$ in flavor space,
where we order the flavors as $(u,d,s)$. We will focus on the
negative charge $Q_e=-Q$, coupling it to the (negative) 
electrostatic potential $\mu_e$.
($\mu_e>0$ corresponds to a density of electrons; $\mu_e<0$
to positrons.)
Enforcing electric neutrality means
choosing $\mu_e$ so that $Q_e$ vanishes.

In a color superconducting phase, there is an expectation
value for some diquark operator. Since a diquark cannot
be electrically neutral and cannot be color neutral,
some subgroup of $U(1)_3 \times U(1)_8 \times U(1)_Q$ 
is spontaneously broken. As we shall see, however,
there is typically at least one linear combination of $T_3$,
$T_8$ and $Q$ with respect to which the condensate 
is neutral, meaning that there is at least one $U(1)$ 
subgroup of $U(1)_3 \times U(1)_8 \times U(1)_Q$ that
the condensate leaves unbroken. We now discuss 
unbroken and broken gauged charges in turn.

\underline{Unbroken gauged charges}. 
Any uniform phase (we will not consider mixed phases,
although they may be important in some contexts), must be neutral
under unbroken gauged charges, to avoid the infrared-divergent energy
cost of long-range electric fields.  
The corresponding chemical (i.e.~electrostatic) 
potential is forced to the 
value of $\mu_Q$ that solves
\beq
Q = \frac{\p \Om}{\p \mu_Q} = 0.
\eeq
Choosing this value of $\mu_Q$ changes the contribution of
any particles with $Q$-charge to the 
free energy of the system. The ensuing free energy cost is proportional
to the volume, though, whereas the free energy cost of the 
electric fields that would arise if neutrality were not enforced
grows with system size faster than the volume.

A macroscopic but finite chunk of matter, that has $\mu_Q\neq 0$
within it chosen so that it is neutral, has a $Q$-electric 
field ({\it i.e.}~a gradient in $\mu_Q$) across its surface.
This $Q$-electric field corresponds to
a layer of polarization.  (For analogous
surface phenomena occurring at the interface between nuclear and
quark matter, see Ref.~\cite{ARRW}.)  We will not consider
surfaces in this paper.  
For our purposes, it suffices to choose $\mu_Q$
to enforce $Q$-neutrality, thus obtaining matter that
can sensibly be analyzed in the infinite volume limit.

\underline{Broken gauged charges}.
The argument requiring that any uniform phase be neutral applies
to broken symmetries too. If a finite-sized sample has a net
charge then there will be electric fields outside the sample
that grow in strength with the size of the sample. This
means that the sample must be neutral to have a good large-volume limit.

Thus for the purposes of imposing neutrality
we can treat $T_3$, $T_8$ and $Q$ on the same footing,
regardless of which linear combinations are broken and which
are unbroken.

\section{Free energy comparisons}

In this section, we compute the free energy of noninteracting
(and thus unpaired) quark
matter, CFL quark matter, 2SC quark matter, and a 
variant of 2SC quark matter in which $u$ and $s$ quarks
(rather than $u$ and $d$ quarks) pair. 
We work to order $(M_s/\mu)^4$
and $(\De/\mu)^2$.  We shall see that 
the free energies of these different phases 
must be compared in a regime where $\De\sim M_s^2/\mu$, meaning that
an expansion of this sort is formally consistent. 
The analysis would be considerably more difficult
were we to attempt to work to higher order in $M_s$, but 
this will not be necessary to make the qualitative
points we wish to make. Quantitatively,
we are interested in $\mu\sim 500$ MeV 
and it is likely that $120 < M_s < 500~\MeV$ 
while $\De \sim 10$ to $100~\MeV$.

We shall work in unitary gauge, where the color-direction of the
quark pairing is chosen to be position-independent. It is always
possible to make gauge transformations that make the color-direction
of the quarks vary with position and transfer some
color into the off-diagonal glue fields.
This would change nothing, although it would make the
analysis less transparent.

In our analysis, we do not include any contribution to the free energy
from the nonabelian gauge bosons. In principle, these can carry charge and 
so one may worry that they
contribute to maintaining (or upsetting) color
neutrality. In unitary gauge,
the color current carried by gluons arises from the gauge-connection part
of the covariant derivative of the gauge field,
$ J_{\rm glue}^\nu =  g [ A_\mu,F^{\mu\nu} ]$. In order for
the charge $J_{\rm glue}^0$ to be nonzero, there would have to be a 
nonzero color electric field $F^{i0}$, which is impossible in
a conducting or broken (superconducting) phase.

\subsection{Unpaired quark matter}
\label{sec:unpaired}

We begin with noninteracting quark matter.  If there really
were no interactions, and therefore global symmetries only,
there would be no reason for the quark matter
to be neutral in any sense.  So, we imagine turning on arbitrarily
small gauge couplings, as this motivates the requirement
that we impose electric and color neutrality and satisfy
weak equilibrium, and ask what chemical potentials we must
introduce in order to achieve neutrality.  
Given that quark
masses are independent of color, we simply set $\mu_3=\mu_8=0$
and obtain quark matter in which the up, down and strange quarks
are all separately color neutral. Imposing electrical
neutrality is (a little) more challenging.
Weak equilibrium relates the chemical potentials for
the three flavors to just $\mu$ and $\mu_e$:
\begin{eqnarray}
\mu_u &=& \mu - \twothirds \mu_e \nonumber\\
\mu_d &=& \mu + \third \mu_e \nonumber\\
\mu_s &=& \mu + \third \mu_e \ ,
\end{eqnarray}
independent of color because $\mu_3=\mu_8=0$.  The free energy
is minimized by filling Fermi seas for each quark flavor, up to Fermi
momenta given by:
\begin{eqnarray}
p_F^u &=& \mu_u \nonumber\\
p_F^d &=& \mu_d  \nonumber\\
p_F^s &=& \sqrt{\mu_s^2 - M_s^2} \ ,
\end{eqnarray}
since the energy of a 
strange quark with momentum $p_F$ is $\sqrt{p_F^2 + M_s^2}$.
The free energy of unpaired quark matter is then given by
\beq\label{OmegaUnpaired}
\ba{rcl}
\Omega_{\rm unpaired}(\mu,\mu_e,M_s) &=& 
 \dsp \frac{3}{\pi^2}\int_0^{p_F^u}
 (p-\mu_u)p^2 dp + \frac{3}{\pi^2}\int_0^{p_F^d}
 (p-\mu_d)p^2 dp \nonumber\\[2ex]
&+& \dsp \frac{3}{\pi^2}\int_0^{p_F^s}
(\sqrt{p^2+M_s^2}-\mu_s)p^2 dp 
+ \frac{1}{\pi^2}\int_0^{\mu_e}(p-\mu_e)p^2 dp\ .
\ea
\eeq
The last term, which is just 
$-\mu_e^4/12\pi^2$, is the free energy of the electrons that
are present because $\mu_e>0$.
We now fix $\mu_e$ by requiring electrical neutrality
\begin{equation}
\frac{\partial \Omega_{\rm unpaired}}{\partial \mu_e}=0 \ .
\label{elecneutralitydef}
\end{equation}
To lowest nontrivial order in $M_s$, this yields
\begin{equation}
\mu_e=\frac{M_s^2}{4\mu}\ ,
\label{mueUnpaired}
\end{equation}
meaning that in electrically neutral noninteracting quark matter
the Fermi momenta are given by 
\begin{eqnarray}
p_F^d &=& \mu + \frac{M_s^2}{12\mu} = p_F^u + \frac{M_s^2}{4\mu} \nonumber\\
p_F^u &=& \mu - \frac{M_s^2}{6\mu} \nonumber\\
p_F^s &=& \mu - \frac{5 M_s^2}{12\mu}= p_F^u - \frac{M_s^2}{4\mu} \ ,
\label{UnpairedFermimomenta}
\end{eqnarray}
to lowest order.  Once (\ref{elecneutralitydef}) is satisfied, $\mu_e$
does not occur linearly in $\Omega$. This means that to see the
first effects of $\mu_e$, we need to work to order $\mu_e^2\sim M_s^4$.
To this order, upon substituting (\ref{mueUnpaired}) into 
(\ref{OmegaUnpaired}) we find
\begin{equation}
\Omega_\unp^\neut = - \frac{3 \mu^4}{4\pi^2} 
+ \frac{3 M_s^2 \mu^2}{4\pi^2} 
- \frac{7- 12\log(M_s/2\mu)}{32\pi^2}M_s^4\ .
\label{Unpairedresult}
\end{equation}

Several points are worth noting before we proceed:
\begin{itemize}
\item
The electronic contribution to $\Omega$ is of order $\mu_e^4\sim M_s^8$,
and therefore does not appear in (\ref{Unpairedresult}).
We therefore neglect it from the beginning
in our analyses of paired phases below. 
%But, in any of the 
%phases we discuss below in which $\mu_e\neq 0$, electrons are present
%and affect $\Omega$ at order $M_s^8$.  
\item
Below, we will mostly ignore interactions, but we will
include the effects of BCS pairing, which gives rise to
gaps $\Delta$ for some or all of the quarks.
Interactions of course result
in other changes to the free energy. 
For example, the
perturbative (in the QCD coupling constant)
corrections to the free energy are 
well-studied~\cite{FreedmanMcLerran}.
The leading effect is a change in the coefficient of
the $\mu^4$ term in the free energy.  
These effects are the same
for all the phases of quark matter we consider, and thus cancel
in the comparisons we shall make. (More precisely, any differences
between the free energies of the different color superconducting
phases introduced by
perturbative QCD interactions 
are perturbative corrections to the $\Delta^2\mu^2$ effects we 
consider explicitly.) These effects would matter only if 
we compared the
free energy of any of the quark matter phases we analyze
to that of hadronic matter.  
Similarly, we leave the bag constant
out since it also only matters
in comparisons between hadronic and quark matter.
We shall not attempt such comparisons.
\item
Note that
the effect of the strange quark mass, combined with the
requirement of electric neutrality, is to push $p_F^d$
up and $p_F^s$ down relative to $p_F^u$.   It is clear
that, at some point, this will cause the CFL phase in
which all flavors pair with each other to be disfavored
relative to unpaired quark matter. The question we
seek to answer is whether at that point pairing of up
with down remains possible.
\end{itemize}

\subsection{Color-flavor locked quark matter}
\label{sec:CFL}

In color-flavor locked quark matter~\cite{CFL},  quarks 
form a condensate in which
\begin{equation}
\langle q^\alpha_a C \gamma_5 q^\beta_b \rangle \sim 
\Delta \left( \epsilon^{\alpha\beta 1}\epsilon_{ab1} 
+ \epsilon^{\alpha\beta 2}\epsilon_{ab2} + 
\epsilon^{\alpha\beta 3}\epsilon_{ab3} \right)\ ,
\label{CFLcondensate}
\end{equation}
where $\alpha,\beta$ are color indices $(r,g,b)$ and $a,b$ are flavor
indices $(u,d,s)$. This form follows from the QCD interaction,
which is attractive in the color-antisymmetric channel, and from
requiring that rotational invariance be preserved, which leads to
a spin-antisymmetric state. The flavor structure is then forced to
be antisymmetric.\footnote{We neglect the
additional condensate which is symmetric
in color and flavor, because although it must be nonzero
it is much smaller than 
the antisymmetric
condensate (\ref{CFLcondensate})~\cite{CFL}, and
because its presence does not modify the symmetries of
CFL quark matter~\cite{ABR2+1}.}
Each of the three terms in (\ref{CFLcondensate}) satisfy these
antisymmetry requirements; summing the three of them allows
all nine quarks to pair, maximizing the pairing energy.
In this state, $rd$ and $gu$ quarks pair yielding two
quasiparticles with gap $\Delta$,  $rs$ and $bu$ quarks pair yielding two
quasiparticles with gap $\Delta$, $bd$ and $gs$ quarks pair yielding two
quasiparticles with gap $\Delta$, and $ru$, $gd$ and $bs$
quarks pair yielding two quasiparticles with gap $\Delta$
and one with gap $2\Delta$.  The $U(1)_3\times U(1)_8\times U(1)_Q$
symmetry
is broken to $U(1)_{\tilde Q}$, where $\tilde Q = Q - T_3 -\half T_8$
is the generator of the unbroken symmetry. 

To order $\Delta^2$, the free energy of the CFL phase can
be described as follows.  One begins with the
free energy of a (fictional) state of unpaired quark matter
in which all quarks that are ``going to pair'' have 
a common Fermi momentum $p_F^{\rm common}$, with $p_F^{\rm common}$
chosen to minimize the free energy of this fictional
unpaired state.  The binding energy of
the diquark condensate is included by subtracting $\Delta^2\mu^2/4\pi^2$ 
for every quasiparticle with gap $\Delta$.
For all the phases we study, $\mu_e, \mu_3, \mu_8$ are of order $M_s^2/\mu$,
so terms such as $\mu_e\mu\De^2$, $\mu_e^2\De^2$, 
etc, can be consistently neglected.
Thus,
\begin{equation}
\Omega_{\rm CFL}= \frac{1}{\pi^2}\sum_{i=1}^{9}\int_0^{p_{F,i}^{\rm common}}
\Bigl(\sqrt{p^2+M_i^2}-\mu_i\Bigr)p^2 dp  - \frac{3}{\pi^2} \Delta^2 \mu^2\ ,
\label{OmegaCFL}
\end{equation}
where the sum runs over all nine quarks, with $M_i=0$ for
the up and down quarks and $M_i=M_s$ for the strange quarks.
The chemical potential
$\mu_i=\mu - Q\mu_e + T_3 \mu_3 + T_8 \mu_8$ for each quark
is specified by its electric and color charges.
Quarks that pair with each other have the same $p_{F,i}^{\rm common}$:
it is in this sense that the parameter is ``common''.
Since not all nine quarks pair with each other,
it appears that there can be four different
values of $p_{F,i}^{\rm common}$ for $(ru,gd,bs)$, $(rd,gu)$, $(rs,bu)$,
and $(gs,bd)$, respectively.\footnote{We thank Sanjay Reddy for
pointing this out.} 
Each of these four parameters should be
determined by minimizing the free energy with respect to it.
We shall see below that in neutral CFL quark matter,
it turns out that all the $p_{F,i}^{\rm common}$ take
on the same value.
This form for the free energy of a state with BCS pairing
between different species goes back to the work of
Clogston and Chandrasekhar \cite{Clogston}, and was derived
in the CFL context in Refs. \cite{neutrality,ARRW}.
Note that it is only valid to order $\Delta^2$ and $M_s^4$,
and that it follows from substituting back into a more general
expression the value of $\De$ that
solves the gap equation; in other words it 
has already been minimized with respect to
$\De$~\cite{RajWilReview}.

We emphasize that, to this order, the nature of the interaction
that generates $\Delta$ does not matter.  The free energy
is given by this prescription regardless of whether the
pairing is due to a point-like four-fermion interaction,
as in NJL models, or due to the exchange of a gluon, as in
QCD at asymptotically high 
energies~\cite{MSW,SchaeferPatterns,RajWilReview}.  
Of course, the strength
and form of the interaction determine the value of $\Delta$, which
we shall keep as a free parameter.

For massless quarks, evaluating $p_F^{\rm common}$
yields simply the average of the chemical potentials of the
quarks that pair.  Evaluation of $p_F^{\rm common}$
for pairing between massless and strange quarks 
yields the result that
the lowest order effect of $M_s$ 
on $p_F^{\rm common}$ is 
to weight the strange quarks in
the average as if their chemical potential were depressed
by $M_s^2/2\mu$.  
That is, evaluating the parameters $p_{F,i}^{\rm common}$
by minimizing the free energy 
%of fictional unpaired
%quark matter (in which quarks that pair have common Fermi
%momenta) 
with respect to them yields
\begin{eqnarray}
p^{\rm common}_{F,(ru,gd,bs)}&=& \mu-\frac{M_s^2}{6\mu}\nonumber\\
p^{\rm common}_{F,(rd,gu)}&=& \mu-\sixth\mu_e+\third\mu_8\nonumber\\
p^{\rm common}_{F,(rs,bu)}&=& \mu-\sixth\mu_e+\quarter\mu_3-\sixth\mu_8
-\frac{M_s^2}{4\mu}\nonumber\\
p^{\rm common}_{F,(gs,bd)}&=& \mu+\third\mu_e-\quarter\mu_3-\sixth\mu_8
-\frac{M_s^2}{4\mu}\ ,
\label{fourpFcommons}
\end{eqnarray}
to lowest order.
This is sufficient
to obtain $\Omega_{\rm CFL}$ to order $M_s^4$.
Note that the quark number densities in the CFL phase are
{\it not} simply given by
terms proportional to $(p_F^{\rm common})^3$:
$\partial\Omega_{\rm CFL}/\partial \mu$ receives a contribution
also from the $\Delta^2\mu^2$ term.  And, of course, in a paired
state there is no sharp Fermi surface anyway as the single particle
density of states is smeared out by of order $\Delta$.
Thus the fictional unpaired quark matter used in the 
construction of $\Omega_{\rm CFL}$ is fictional in two senses:
(i) the Fermi momenta are {\it not} the same as
in the unpaired quark matter of Section \ref{sec:unpaired};
(ii) the CFL state has no Fermi momentum and
the parameters $p_{F,i}^{\rm common}$ that
arise in its description do not fully specify the quark number
density in the CFL state.

\noindent\parbox{\textwidth}{
\ \ \ \ To the order at which we are working, 
\begin{eqnarray}
\Omega_{\rm CFL} = - \frac{3 \mu^4}{4\pi^2} 
&+& \frac{3 M_s^2 \mu^2}{4\pi^2}
-\frac{1}{96\pi^2}\Biggl[ \left( 16 \mu_e^2 + 12 \mu_3^2 + 16 \mu_8^2
- 24 \mu_3 \mu_e - 16 \mu_e\mu_8\right)\mu^2\nonumber\\
&+& \left( 16 \mu_8  - 8 \mu_e \right) M_s^2 \mu 
+7 M_s^4 - 36 M_s^4 \log(M_s/2\mu) \Biggr]
- \frac{3\Delta^2\mu^2}{\pi^2}\ .\nonumber\\
~&~&
\label{CFLOmega}
\end{eqnarray}
}
Hence,
\begin{equation}
\ba{rcl}
\dsp\frac{\partial \Omega_{\rm CFL}}{\partial \mu_e} &=&
  \dsp \frac{1}{12\pi^2}\Bigl(
  M_s^2 \mu + 3 \mu_3\mu^2 + 2\mu_8\mu^2 - 4\mu_e\mu^2 \Bigr)
  \\[2ex]
\dsp\frac{\partial \Omega_{\rm CFL}}{\partial \mu_3} &=& 
 \dsp \frac{\mu^2}{4\pi^2} ( \mu_e-\mu_3)
 \\[2ex]
\dsp\frac{\partial \Omega_{\rm CFL}}{\partial \mu_8} &=&
  \dsp \frac{1}{6\pi^2}\Bigl(
  -M_s^2 \mu - 2\mu_8\mu^2 + \mu_e\mu^2 \Bigr)\ .
\ea
\label{CFLneutralityconditions}
\end{equation}
In discussing the neutrality of the CFL phase, it is 
convenient to use the basis of broken ($X,Y$) and unbroken ($\Qt$)
charges,
\beq
\ba{rcl@{\qquad}rcl}
\Qt &=& Q - T_3 - \half T_8 & 
 \mu_\Qt &=& \frac{4}{9}( \mu_Q - \mu_3 - \half \mu_8)  \\[1ex]
X &=& -Q + T_3 - 4 T_8 & 
 \mu_X &=& \frac{1}{18}(-\mu_Q + \mu_3 - 4 \mu_8)  \\[1ex]
Y &=& -Q - T_3 &  
 \mu_Y &=& \half (-\mu_Q - \mu_3) \ ,
\ea
\eeq
where $\mu_Q=-\mu_e$.
Then, we see from (\ref{CFLneutralityconditions}) 
that, to the order we are working, electric
and color neutrality require 
\begin{eqnarray}
\mu_X &=&  \frac{M_s^2}{9\mu}\nonumber\\
\mu_Y &=& 0\ .
%\nonumber\\
%\mu_\Qt &=& {\rm anything}.
\end{eqnarray}
The role of the nonzero $\mu_X$ is to enforce $X$-neutrality,
and it does so by forcing the apparently 
different $p^{\rm common}_{F,i}$'s in (\ref{fourpFcommons})
to all be
equal, as was assumed without proof in Ref.~\cite{ARRW}.
To this order, we see from (\ref{CFLneutralityconditions}) 
that $\Omega_{\rm CFL}$
does not depend on 
$\mu_{\tilde Q}$, so that this combination of chemical potentials
cannot be specified by the requirement of neutrality.
This means that CFL quark matter is a $\tilde Q$-insulator~\cite{neutrality}.
As we discuss below, this result holds even at higher order.
Even though CFL quark matter is a $\tilde Q$-insulator, 
meaning that it is neutral for a range of values of $\mu_\Qt$,
in order to obtain neutral matter we must ensure that,
once we go to higher order, no electrons are present. 
Electrons give contributions to
$\Om$ of the form $\mu_e^4 = (-\mu_\Qt +\mu_X +\mu_Y)^4$,
and enforcing neutrality 
therefore fixes $\mu_\Qt = \mu_X$.\footnote{An electron quasiparticle
excitation above the CFL ground
state should be thought of as carrying $\Qt$ charge. This can
be described by saying that if an electron is injected into a lump
of CFL material from outside it, the $X$-charge of the electron
will be screened by the CFL condensate, leaving the electron
excitation propagating within the CFL matter carrying only
$\Qt$-charge. As this example makes clear, though, in order to
enforce electrical neutrality it is the total $Q$-charge
of the electron that matters, regardless of the fact that 
the electron carries its $\Qt$ with it while losing its $X$ 
to the condensate.  Thus, when we enforce neutrality at
higher order by ensuring that no electrons are present,
we do so by setting
$\mu_e=0$ and thus $\mu_\Qt = \mu_X\neq 0$.}

To order $M_s^4$, the free energy for electric and color neutral CFL quark
matter is given by \cite{ARRW}
\begin{eqnarray}
\Omega_{\rm CFL}^\neut&=& - \frac{3 \mu^4}{4\pi^2} 
+ \frac{3 M_s^2 \mu^2}{4\pi^2} 
- \frac{1- 12\log(M_s/2\mu)}{32\pi^2}M_s^4 
- \frac{3\Delta^2\mu^2}{\pi^2}\nonumber\\
&=& \Omega_{\rm unpaired}^\neut + \frac{ 3 M_s^4 - 48\Delta^2\mu^2}{16\pi^2}\ ,
\end{eqnarray}
We conclude that CFL quark matter is favored over unpaired
quark matter as long as the interactions are strong enough
to generate a gap in the CFL phase satisfying
\begin{equation}
\Delta > \frac{M_s^2}{4\mu}\ .
\label{CFL-unpaired-criterion}
\end{equation}
Since $\Delta$ is of order $M_s^2$ in this criterion, our 
strategy of evaluating $\Omega$ to order $M_s^4$ and 
to order $\Delta^2$ is 
consistent.

In the calculation above, 
we have neglected the fact that when $M_s\neq 0$,
the gaps in the CFL phase are $M_s$-dependent and, furthermore,
are non-degenerate \cite{ABR2+1,SW2}.  For example, the gap for
the quasiparticles obtained upon pairing the
$rd$ and $gu$ quarks differs from that 
obtained upon pairing the  $rs$ and $bu$ quarks.
The leading $M_s$ dependence of the latter is 
$\Delta(M_s)\sim \Delta(0)(1 - c M_s^2/\mu^2)$, for 
some coefficient~$c$~\cite{massloff}. These effects are therefore
of order $\Delta(0)^2 M_s^2$ in the free energy and
it is consistent to neglect them.  Although
their neglect is consistent, they are 
important because they introduce new 
$\mu_X$ and $\mu_Y$ dependence into the free energy.
At a high enough order that the differences among
gaps matters, the fact that different Cooper pairs
in the CFL phase have different $X$ and $Y$ 
charges matters. (At this order, the
expression (\ref{OmegaCFL}) is itself incomplete.)
It is worth noting that the $M_s$-dependent
differences among gaps cannot introduce any dependence on 
$\mu_{\tilde Q}$ because, by
the definition of $\tilde Q$, each Cooper pair
in the CFL phase has $\tilde Q=0$. (For example, $rd$ with
$\tilde Q=-1$ pairs only with $gu$ with $\tilde Q=+1$.)
The CFL phase is $\tilde Q$-neutral
and a $\tilde Q$-insulator, even at higher order
in $M_s$.
On the other hand, CFL Cooper pairs do have $X$ and $Y$
charges. The total $X$ and $Y$ charge of the condensate
vanishes only to the extent that all the gaps are the same.
This means that
imposing neutrality with respect to the broken gauge charges
of the CFL phase will be less trivial at higher order
than it is at the order we are working. This warrants investigation.

We have also neglected the possibility that when $M_s\neq 0$,
the left-handed CFL condensate and the right-handed CFL condensate
may rotate relative to one another in flavor 
space~\cite{BedaqueSchaefer,KaplanReddy}.
By neglecting this kaon condensation, we have neglected an effect
that, if it occurs, lowers the free energy of the CFL phase 
at order $M_s^4$~\cite{BedaqueSchaefer} meaning that in
this instance our neglect is not consistent.  If we
take the quantitative results for the correction
to $\Omega_{\rm CFL}$ due to maximal kaon
condensation given in Ref.~\cite{BedaqueSchaefer}, we find that
the $4$ in the denominator of (\ref{CFL-unpaired-criterion})
increases but remains smaller than $5$.  
We shall see below that in the comparison
between CFL and 2SC, there is a much larger uncertainty to 
be concerned with.

\subsection{2SC quark matter}

In 2SC quark matter,  quarks 
form a condensate in which
\begin{equation}
\langle q^\alpha_a C \gamma_5 q^\beta_b \rangle \sim 
\Delta \epsilon^{\alpha\beta 3}\epsilon_{ab3} \ .
\label{2SCcondensate}
\end{equation}
In this state, $rd$ and $gu$ quarks pair yielding two
quasiparticles with gap $\Delta$,  and $ru$ and $gd$ quarks
pair, yielding two quasiparticles with gap $\Delta$. 
Five quarks are left unpaired.
The $U(1)_3$ symmetry (and indeed
a color $SU(2)$ symmetry of which it is a subgroup) is left
unbroken, as is 
$U(1)_{\tilde Q}$, where $\tilde Q = Q - \half T_8$.

To the order we are working, the free energy of the 2SC phase 
is
\begin{equation}
\ba{rcl}
\Omega_{\rm 2SC} &=& 
 \dsp \frac{1}{\pi^2}\sum_{i=1}^{4}\int_0^{p_F^{\rm common}}
 (p-\mu_i)p^2 dp  \\[2ex]
&+&
  \dsp \frac{1}{\pi^2}\sum_{i=5}^{9}\int_0^{\sqrt{\mu_i^2-M_i^2}}
 (\sqrt{p^2+M_i^2}-\mu_i)p^2 dp \\
&-& \dsp \frac{1}{\pi^2} \Delta^2 \mu^2\ ,
\ea
\label{Omega2SC}
\end{equation}
where the first four quarks are those that pair while the last
five are those that don't. 
This time, $p_F^{\rm common}$
is just the average of the chemical potentials of the
quarks that pair:
\begin{equation}
p_F^{\rm common}=\mu- \sixth\mu_e+\third\mu_8 \ .
\end{equation}
To the order at which we are working, 
\beq
\ba{rcl}
\dsp \frac{\partial \Omega_{\rm 2SC}}{\partial \mu_e} &=& 
  \dsp \frac{M_s^2\mu - 2\mu_e \mu^2}{2\pi^2}\ ,\\[2ex]
\dsp \frac{\partial \Omega_{\rm 2SC}}{\partial \mu_3} &=& 
  \dsp -\frac{\mu_3\mu^2}{2\pi^2}\ , \\[2ex]
\dsp \frac{\partial \Omega_{\rm 2SC}}{\partial \mu_8}&=& 
  \dsp-\frac{2\mu_8\mu^2}{\pi^2}\ ,
\ea
\eeq
meaning that we find electric and color neutrality with
$\mu_3=\mu_8=0$ and $\mu_e=M_s^2/2\mu$.
Note that $Q$ is a linear combination of the unbroken charge $\tilde Q$
and the orthogonal broken charge $X\equiv T_8-\half Q$. 
It is more natural to describe the 2SC phase in terms of $\Qt$ and
$X$ charges, in which case the neutrality condition becomes
$\mu_{\tilde Q}=-2M_s^2/(5\mu)$ and $\mu_X=-M_s^2/(5\mu)$.
The 2SC phase contains ungapped  $\tilde Q$-carrying modes,
and is therefore a $\tilde Q$-conductor.

To order $M_s^4$, the
free energy for electric and color neutral 2SC quark
matter is given by 
\begin{eqnarray}
\Omega_{\rm 2SC}^\neut &=& - \frac{3 \mu^4}{4\pi^2} 
+ \frac{3 M_s^2 \mu^2}{4\pi^2} 
- \frac{5 - 12\log(M_s/2\mu)}{32\pi^2}M_s^4 
- \frac{\Delta^2\mu^2}{\pi^2}\nonumber\\
&=& \Omega_{\rm unpaired}^\neut + \frac{ M_s^4 - 16 \Delta^2\mu^2}{16\pi^2}\ .
\label{2SCresult}
\end{eqnarray}

\subsection{2SCus quark matter}

There is an obvious variant of the 2SC phase which we 
must consider, before we make the comparison between 2SC and CFL.
In the 2SC phase above, it is the $u$ and $d$ quarks that pair.
Given (\ref{UnpairedFermimomenta}), it seems as likely for $u$ and
$s$ to  pair.  In the resulting 2SCus quark matter,  quarks 
form a condensate in which
\begin{equation}
\langle q^\alpha_a C \gamma_5 q^\beta_b \rangle \sim 
\Delta \epsilon^{\alpha\beta 2}\epsilon_{ab2} \ .
\label{2SCuscondensate}
\end{equation}
In this state, $rs$ and $bu$ quarks pair yielding two
quasiparticles with gap $\Delta$,  and $ru$ and $bs$ quarks
pair, yielding two quasiparticles with gap $\Delta$. 
Five quarks are left unpaired.
The color $U(1)$ symmetry generated by 
$\diag(\half,0,-\half)=\half T_3+\threequarters T_8$ 
(and indeed
a color $SU(2)$ symmetry of which this is a subgroup) is left
unbroken, as is 
$U(1)_{\tilde Q}$, where $\tilde Q = Q - T_3 - \half T_8$.

To the order we are working, the free energy of the 2SCus phase 
is 
\begin{eqnarray}
\Omega_{\rm 2SCus}&=&\frac{1}{\pi^2}\sum_{i=1}^{4}\int_0^{p_F^{\rm common}}
(\sqrt{p^2+M_i^2}-\mu_i)p^2 dp  \nonumber\\
&+&\frac{1}{\pi^2} \sum_{i=5}^{9}\int_0^{\sqrt{\mu_i^2-M_i^2}}
(\sqrt{p^2+M_i^2}-\mu_i)p^2 dp
- \frac{1}{\pi^2} \Delta^2 \mu^2\ ,
\label{Omega2SCus}
\end{eqnarray}
where the first four quarks are those that pair while the last
five are those that don't. 
This time, 
\begin{equation}
p_F^{\rm common}=\mu -\sixth\mu_e+\third\mu_8-\frac{M_s^2}{4\mu}\ .
\end{equation}
To the order at which we are working, 
\begin{eqnarray}
\frac{\partial \Omega_{\rm 2SCus}}{\partial \mu_e} &=& 
  -\frac{\mu_e\mu^2}{\pi^2} \nonumber\\
\frac{\partial \Omega_{\rm 2SCus}}{\partial \mu_3} &=& 
  -\frac{\mu_3\mu^2}{2\pi^2}
\nonumber\\
\frac{\partial \Omega_{\rm 2SCus}}{\partial \mu_8}&=& 
  -\frac{2\mu_8\mu^2}{\pi^2}\ ,
\end{eqnarray}
meaning that we find electric and color neutrality with
$\mu_3=\mu_8=\mu_e=0$. (At higher order, all would become nonzero.)
The 2SCus phase is a $\tilde Q$-conductor.

To order $M_s^4$, the
free energy for electric and color neutral 2SCus quark
matter is identical to that for electric and color neutral 2SC
quark matter, given in (\ref{2SCresult}).

\subsection{2SC vs. CFL Comparison}

\begin{figure}[t]
\begin{center}
\includegraphics[width=0.6\textwidth,angle=-90]{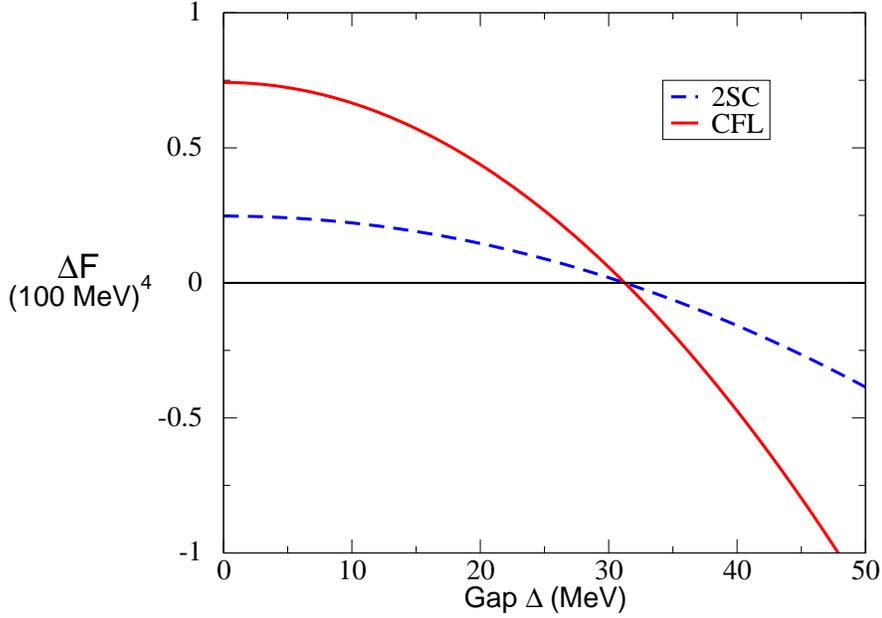}
\end{center}
\caption{
The free energy difference $\De F$ between 
various phases and unpaired quark matter,
as a function of the gap $\De$. The unpaired phase is therefore
the $\De F=0$ horizontal line.
We show the CFL
phase (red/solid) and the 2SC or 2SCus phase (blue/dashed). 
Below about $32~\MeV$ the unpaired phase
is favored, above it CFL is favored. There is
no 2SC window.
We have taken $\mu=500$ MeV and $M_s=250$ MeV.  
If we assume that $\Delta$ and $M_s$ do not change
with $\mu$, then the horizontal axis could equally
well be $\mu$ at fixed $\Delta$ and $M_s$, or
$1/M_s^2$ at fixed $\Delta$ and $\mu$, instead of
$\Delta$ at fixed $\mu$ and $M_s$ as plotted.
The curves cross where $M_s^2/\mu\Delta=4$.
}
\label{fig:OmegasvsDelta}
\end{figure}
The first comparison to make is that between all the phases
we have discussed (unpaired, CFL, 2SC, 2SCus) 
upon making the assumption that $M_s$ is the same in all phases and
$\Delta$ is the same in all the paired phases. This comparison is 
shown in 
Fig.~\ref{fig:OmegasvsDelta}.

If Fig.~\ref{fig:OmegasvsDelta} were the end of the story,
we would conclude that for $\Delta<M_s^2/4\mu$ unpaired quark
matter is favored, whereas for $\Delta>M_s^2/4\mu$ CFL quark
matter is favored. Precisely at $\Delta=M_s^2/4\mu$, unpaired, CFL,
2SC and 2SCus quark matter all have the same free energy. There
is therefore (just barely) no 2SC window.  Given the fragility
of this conclusion, let us enumerate the effects we know
we have left out:
\begin{itemize}
\item
Effects that are of order $M_s^6$ and higher.
We have discussed some of these above.  We do not
know how they all add up, but we can consistently neglect them.
\item
As mentioned above, we have neglected the possibility of kaon
condensation in CFL quark matter. This can only lower the CFL free
energy. It therefore works against opening
a 2SC window.
\item
For a given interaction, the gap $\Delta$ is {\it not} the same
in the 2SC and CFL phases.  If the interaction is single-gluon
exchange as at asymptotically high densities, the CFL gap
is smaller than the 2SC gap by a factor of 
$2^{-1/3}\approx 0.79$ \cite{SchaeferPatterns}.
If the interaction is a point-like
interaction with the quantum numbers of single-gluon
exchange, the ratio of the CFL gap to the 2SC gap
is $\approx 0.75$ \cite{ABR2+1,RajWilReview}.
The fact that the 2SC gap is somewhat bigger than the CFL gap
tends to open up a 2SC window. This effect is comparable
in magnitude (and opposite in sign) to that of kaon condensation.
\item
The biggest effect comes from the fact that $M_s$ is not the
same in the CFL and 2SC phases.  The free energy
of all the phases we have discussed includes a common 
$3/(4\pi^2)\ M_s^2 \mu^2$ term, which
plays no role in the comparison of Fig.~\ref{fig:OmegasvsDelta}.
However when comparing phases with different $M_s$ it will
parametrically dominate  the $M_s^4$ and $\Delta^2$
effects on which we have focussed to this point. We now discuss
this effect in more detail.
\ei

\begin{figure}[t]
\begin{center}
\includegraphics[width=0.6\textwidth,angle=-90]{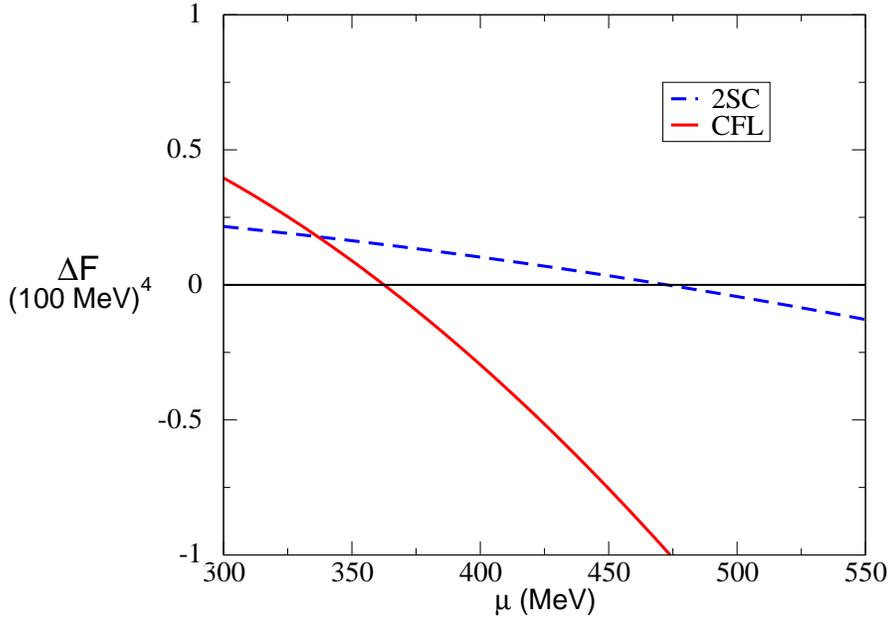}
\end{center}
\caption{
The free energy of the 2SC (blue/dashed) and CFL (red/solid)
phases as a function of the 
quark chemical potential $\mu$,
relative to the unpaired phase with $M_s=275$ MeV (horizontal black line).
We have taken $M_s=250$ MeV and $\Delta=30$ MeV in the CFL
phase, and $M_s=275$ MeV and $\Delta=40$ MeV in the 2SC phase.
These parameters are completely ad hoc, but the 
resulting curves demonstrate
quite clearly that the effect of the difference between $\Delta$'s,
which seeks to open a 2SC window, is easily overpowered
by the effect of even a 
fairly small difference between $M_s$'s, which slams
the window shut.
}
\label{fig:CFLvs2SC}
\end{figure}

A full treatment of how the constituent mass $M_s$ and gaps $\Delta$
vary between different phases would require analysis of the coupled
gap equations in some model Hamiltonian, an exercise that goes beyond
the general considerations we employ in this paper.  
Such an analysis has
recently been done for several NJL models by Buballa and
Oertel~\cite{BuballaOertel}, albeit without enforcing electric
and color neutrality.  They find that $M_s$ is considerably
smaller in the CFL phase than in the 2SC phase.  They do not consider
the 2SCus phase, but it is reasonable to assume that $M_s$ in this
phase will be comparable to that in the 2SC phase.

In Fig.~\ref{fig:CFLvs2SC}, we illustrate the effect of a relatively
small difference in the value of $M_s$ between phases.  We plot the
relative free energy of the CFL, 2SC, and unpaired phases, taking
($M_s=250~\MeV$, $\De=30~\MeV$) in the CFL phase and 
($M_s=275~\MeV$, $\De=40~\MeV$) in the unpaired and 2SC phases.
(In the models of
Ref.~\cite{BuballaOertel}, the change in $M_s$ between the CFL and 2SC
phases is significantly larger, of order $30\%$, and there is no
analogue of our unpaired quark matter.)  We see in
Fig.~\ref{fig:CFLvs2SC} that the fragility of the conclusion of
Fig.~\ref{fig:OmegasvsDelta} is illusory: the 2SC phase is firmly
excluded if $M_s$ varies as Buballa and Oertel find. Because it is an
order $M_s^2$ effect, the greatest uncertainty in our conclusions
arises from our uncertainty in how much $M_s$ changes between
different phases. Our analysis has also neglected $\mu$-independent
contributions to the
free energy that arise from the binding energy of the
$\<\sbar s\>$ condensate, and so favor the 2SC phase where $M_s$ is
larger.  These are subleading relative to the $\mu^2 M_s^2$ term.

\section{Conclusions}
\label{sec:conclusions}

We find that, taking into account the constraints imposed by color
neutrality, electrical neutrality, and weak equilibrium, the
two-flavor color superconducting (2SC) phase is not expected to be
found in compact stars.  Our argument is model independent, and is
based on an expansion in $M_s/\mu$ and $\De/\mu$. Since it is 
possible that in lower density quark matter $M_s$ is not
much smaller than $\mu$, it would
be of considerable interest to perform the full calculation for a
specific model, solving coupled gap equations under the constraint of
electrical and color neutrality, to see whether the 2SC phase vanishes
as expected. An NJL model such as that of Buballa and Oertel
\cite{BuballaOertel} would be a natural
starting point.

Our results indicate that two qualitative
possibilities remain viable for the phase diagram of quark
matter that is electrically and color neutral.  The first 
is a single phase transition directly from hadronic matter to
color-flavor locked quark matter.  The second possibility
is that there may be a window of 
intermediate densities in which we find quark
matter that is, at the level of the analysis of this paper,
unpaired.  That is, as a function of increasing $\mu$
we first go from hadronic matter to unpaired quark matter
and only at a higher $\mu$ make the transition to CFL quark matter.

If there is a window of ``unpaired'' three-flavor quark matter, there
will certainly be pairing in this window:
all that we have shown is that there
will be no BCS pairing between quarks of different flavors.
One possible pattern of pairing is the formation of $\langle uu \rangle$,
$\langle dd \rangle$ and $\langle ss \rangle$ 
condensates. These must be either $J=1$ or symmetric in color,
and are therefore much smaller than the $J=0$ color-antisymmetric
condensates we have investigated in this paper. The gaps 
in these phases may be as large
as of order 1 MeV \cite{Schaefer1Flavor}, or could be much 
smaller \cite{ARW1}.
In all the condensates we have discussed to this point, the
Cooper pairs are made of quarks with equal and opposite
momenta.  Another possibility for pairing in the ``unpaired''
quark matter is crystalline color 
superconductivity \cite{massloff,BowersLOFF,ngloff,pertloff},
which involves pairing between quarks whose momenta do not
add to zero, as first considered in a condensed
matter physics context in Refs. \cite{LOFF}.  
The unpaired quark matter of (\ref{UnpairedFermimomenta})
is susceptible to the formation of a crystalline
color superconducting condensate constructed from pairs
of quarks with different flavors,
both of which have momenta near their
respective unpaired Fermi surfaces.  Our 
work demonstrates that the electrically neutral unpaired quark matter
of Section 3.1 is the correct starting point for an analysis
of crystalline color superconductivity.  
Crystalline color superconductivity
need not coexist with the 2SC phase, as previously thought.

\vspace{3ex}
{\samepage 
\begin{center} {\bf Acknowledgements} \end{center}
\nopagebreak
We have had many illuminating discussions with Sanjay Reddy. 
We thank him for pointing
out an error in the derivation of $\Omega_{\rm CFL}^{\rm neutral}$
in an earlier draft.
KR is grateful for the support and hospitality of the Institute
for Theoretical Physics at Santa Barbara.
The research of KR 
is supported in part by the DOE under cooperative research agreement
DE-FC02-94ER40818 and by the NSF under Grant No. PHY99-07949.
The research of MGA is supported in part by
the UK PPARC.
}

\end{document}